\documentclass[12pt]{article}
\usepackage{amsfonts}
\usepackage{bm}
\usepackage{graphicx}
\usepackage{multicol}

\begin{document}

\title{Time reversal and rotational symmetries in noncommutative phase space}
\maketitle

\centerline {Kh. P. Gnatenko \footnote{E-Mail address: khrystyna.gnatenko@gmail.com}, M. I. Samar \footnote{E-Mail address: mykolasamar@gmail.com}, V. M. Tkachuk \footnote{E-Mail address: voltkachuk@gmail.com}}
\medskip

\centerline {\small  $^{1,2,3}$ \it  Ivan Franko National University of Lviv,}
\centerline {\small \it Department for Theoretical Physics,12 Drahomanov St., Lviv, 79005, Ukraine}
\centerline {\small $^{1}$ \it  Laboratory for Statistical Physics of Complex Systems}
\centerline {\small \it  Institute for Condensed Matter Physics, NAS of Ukraine Lviv, 79011, Ukraine}

\begin{abstract}
Time reversal symmetry is studied in a space with noncommutativity of coordinates and noncommutativity of momenta of canonical type. The circular motion is examined as an apparent example of time reversal symmetry breaking in the space. On the basis of exact solution of the problem we show that because of noncommutativity the period of the circular motion depends on its direction. We propose the way to recover the time reversal and rotational symmetries in noncommutative phase space of canonical type. Namely, on the basis of idea of generalization of parameters of noncommutativity to tensors we construct noncommutative algebra which is rotationally-invariant, invariant under time reversal, besides it is equivalent to noncommutative algebra of canonical type.

\end{abstract}

\section{Introduction}

Much attention has been devoted to studies of a quantum space realized on the basis of idea that coordinates can be noncommutative. The idea was suggested by Heisenberg. Later it was formalized by Snyder \cite{Snyder}.
Noncommutative phase space of canonical type has been studied intensively.  It is characterized by the following commutation relations for operators of coordinates and momenta
  \begin{eqnarray}
[X_{i},X_{j}]=i\hbar\theta_{ij},\label{form101}\\{}
[X_{i},P_{j}]=i\hbar(\delta_{ij}+\gamma_{ij}),\label{form1001}\\{}
[P_{i},P_{j}]=i\hbar\eta_{ij},\label{form10001}{}
\end{eqnarray}
with $\theta_{ij}$, $\eta_{ij}$, $\gamma_{ij}$ being elements of constant antisymmetric matrixes.

Idea that coordinates and momenta may be noncommutative gives a possibility to construct quantum space (space with minimal length) at the same time noncommutativity causes fundamental problems. In noncommutative space of canonical type (\ref{form101})-(\ref{form10001}) with $\eta_{ij}=\gamma_{ij}=0$ one face the problems of rotational and the time-reversal symmetries breaking \cite{Chaichian,Balachandran1,Geloun,Scholtz}. The same problems appear in noncommutative phase space (\ref{form101})-(\ref{form10001}).

 To preserve rotational symmetry the noncommutative algebra of canonical type was generalized in different ways as a result  different types of algebras with noncommutativity of coordinates were proposed and examined \cite{Moreno,Galikova,Amorim,GnatenkoPLA14}). Among these algebras rotationally invariant algebra with position-dependent noncommutativity  (see, for example, \cite{Lukierski,Lukierski2009,BorowiecEPL,Borowiec,Borowiec1,Kupriyanov2009,Kupriyanov}), noncommutative algebras with spin noncommutativity  (see, for example, \cite{Falomir09,Ferrari13}) were intensively studied. These algebras are rotationally invariant but they are not equivalent to noncommutative algebras of canonical type in the sense that the relation $[X_i,\theta_{ij}]=[P_i,\theta_{ij}]=0$ does not hold in the frame of the algebras.

 In the present paper we study time reversal symmetry in  noncommutative phase space of canonical type (\ref{form101})-(\ref{form10001}).  The circular motion is studied as an obvious example for observing the time reversal symmetry breaking. On the basis of exact solution of the problem  it is shown that because of noncommutativity the period of the circular motion depends on its direction. So, noncommutativity causes violation of the time reversal symmetry. To recover the symmetry  we propose noncommutative algebra which is  time reversal invariant, rotationally invariant, and equivalent to noncommutative algebra of canonical type.

 The paper is organized as follows. In Section 2 the time reversal symmetry is studied in noncommutative phase space of canonical type. The invariance of noncommutative algebra upon time reversal is analyzed. Influence of noncommutativity on the period of circular motion in different directions is examined. In Section 3 algebra with noncommutativity of coordinates and noncommutativity of momenta which is time reversal and rotationally-invariant is constructed.  Conclusions are presented in Section 4.

\section{Time reversal symmetry in noncommutative phase space of canonical type  }

Obvious example for observing violation of the time reversal symmetry in noncommutative phase space is a circular motion.  The effect of noncommutativity on the motion depends on its direction.
To show this let us consider  noncommutative algebra of canonical type in two-dimensional case
  \begin{eqnarray}
[X_{1},X_{2}]=i\hbar\theta,\label{2dform101}\\{}
[X_{1},P_{1}]=[X_{2},P_{2}]=i\hbar(1+\gamma),\label{2dform1001}\\{}
[P_{1},P_{2}]=i\hbar\eta,\label{2dform10001}{}
\end{eqnarray}
and study the following Hamiltonian
\begin{eqnarray}
H=\frac{P_1^2}{2m}+\frac{P_2^2}{2m}-\frac{k}{X}.\label{hc1}
\end{eqnarray}
Here $\theta$, $\eta$, $\gamma$ being constants, $X=\sqrt{X^2_1+X^2_2}$.
In the classical limit $\hbar\rightarrow0$ from (\ref{2dform101})-(\ref{2dform10001}) we have the following Poisson brackets
\begin{eqnarray}
\{X_{1},X_{2}\}=\theta,\label{ppform101}\\{}
\{X_{1},P_{1}\}=\{X_{2},P_{2}\}=1+\gamma,\label{ppform1001}\\{}
\{P_{1},P_{2}\}=\eta.\label{ppform10001}{}
\end{eqnarray}
Taking into account that $X_i$, $P_i$ in (\ref{hc1}) satisfy (\ref{ppform101})-(\ref{ppform10001}),
one obtains  equations of motion
\begin{eqnarray}
\dot{X}_1=\frac{P_1}{m}\left(1+\gamma\right)+\frac{k\theta X_2}{X^3},\label{m}\\
\dot{X}_2=\frac{P_2}{m}\left(1+\gamma\right)-\frac{k\theta X_1}{X^3},\\
\dot{P}_1=\frac{\eta P_2}{m}-\frac{k X_1}{X^3}\left(1+\gamma\right),\\
\dot{P}_2=-\frac{\eta P_1}{m}-\frac{k X_2}{X^3}\left(1+\gamma\right).\label{mm}
\end{eqnarray}
The obtained equations have solution
\begin{eqnarray}
{X}_1(t)=R_0\cos(\omega t),\ \ {X}_2(t)=R_0\sin(\omega t),\label{sol}\\
{P}_1(t)=-P_0\sin(\omega t),\ \ P_2(t)=P_0\cos(\omega t).\label{sol1}
\end{eqnarray}
The solution corresponds to circular motion with radii $R_0$, momentum
\begin{eqnarray}
P_0=\frac{m\omega R^3_0+k m\theta}{R_0^2\left(1+\gamma\right)},\label{p00}
\end{eqnarray}
and frequency
\begin{eqnarray}
\omega=\frac{1}{2}\left(\sqrt{\frac{4k}{m R_0^3}\left((1+\gamma)^2-\theta\eta\right)+\left(\frac{k\theta}{R^3_0}+\frac{\eta}{m}\right)^2}-\frac{\eta}{m}-\frac{k\theta}{R_0^3}\right).\label{omega}
\end{eqnarray}
The period of the motion reads
\begin{eqnarray}
T=4\pi\left(\sqrt{\frac{4k}{m R_0^3}\left((1+\gamma)^2-\theta\eta\right)+\left(\frac{k\theta}{R^3_0}+\frac{\eta}{m}\right)^2}-\frac{\eta}{m}-\frac{k\theta}{R_0^3}\right)^{-1}.\label{period}
\end{eqnarray}

Let us consider the circular motion in opposite direction with radii $R_0$. The corresponding solution reads
\begin{eqnarray}
{X}_1(t)=R_0\cos(\omega t),\ \ {X}_2(t)=-R_0\sin(\omega t),\label{s}\\
{P}_1(t)=P^{\prime}_0\sin(\omega t),\ \ P_2(t)=P^{\prime}_0\cos(\omega t).\label{ss}
\end{eqnarray}
Expressions (\ref{s}), (\ref{ss}) correspond to (\ref{sol}), (\ref{sol1}) with $-t$. In (\ref{ss}) we put $P^{\prime}_0$ to distinguish momentum which corresponds to the motion in opposite direction.
Substituting (\ref{s}), (\ref{ss}) into (\ref{m})-(\ref{mm}) one obtains
\begin{eqnarray}
\omega^{\prime}=\frac{1}{2}\left(\sqrt{\frac{4k}{m R_0^3}\left((1+\gamma)^2-\theta\eta\right)+\left(\frac{k\theta}{R^3_0}+\frac{\eta}{m}\right)^2}+\frac{\eta}{m}+\frac{k\theta}{R_0^3}\right),\label{omega1}\\
T^{\prime}=4\pi\left(\sqrt{\frac{4k}{m R_0^3}\left((1+\gamma)^2-\theta\eta\right)+\left(\frac{k\theta}{R^3_0}+\frac{\eta}{m}\right)^2}+\frac{\eta}{m}+\frac{k\theta}{R_0^3}\right)^{-1},\label{period1}
\end{eqnarray}
and
 \begin{eqnarray}
P^{\prime}_0=-\frac{m\omega^{\prime} R^3_0-k m\theta}{R_0^2\left(1+\gamma\right)}.\label{pp}
\end{eqnarray}
 Note that the obtained frequency (\ref{omega1}) and period (\ref{period1}) do not coincide with  (\ref{omega}), (\ref{period}). One has
 \begin{eqnarray}
\Delta\omega=\omega^{\prime}-\omega=\frac{\eta}{m}+\frac{k\theta}{R_0^3}.
\end{eqnarray}
 The period and the frequency are different for motions in a circle of radius $R_0$ in different directions.  In comparison to (\ref{omega}), (\ref{period}) expressions for $\omega^{\prime}$, $T^{\prime}$ contain parameters of noncommutativity with opposite signs. One also has that $P_0^{\prime}\neq-P_0$ (in the ordinary space one has that $P^{\prime}_0=-P_0$ which corresponds to the motion in opposite direction). Discrepancy in expressions (\ref{omega}), (\ref{period}) and (\ref{omega1}), (\ref{period1}) is because of non-invariance of noncommutative algebra (\ref{2dform101})-(\ref{2dform10001}) upon time reversal, because of time reversal symmetry  breaking in noncommutative phase space.

Ordinary commutation relations for coordinates and momenta ($\theta=\eta=\gamma=0$) are invariant under the time reversal transformation \cite{Domingos}. In analogy to the ordinary case ($\theta=\eta=\gamma=0$), considering transformations of coordinates and momenta upon time reversal as  $X_i\rightarrow X_i$, $P_i\rightarrow -P_i$  and taking into account that in quantum mechanics the time reversal operation involves complex conjugation \cite{Domingos},  for (\ref{2dform101})-(\ref{2dform10001}) one obtains commutation relations
  \begin{eqnarray}
[X_{1},X_{2}]=-i\hbar\theta,\label{aform101}\\{}
[X_{1},P_{1}]=[X_{2},P_{2}]=i\hbar(1+\gamma),\label{aform1001}\\{}
[P_{1},P_{2}]=-i\hbar\eta,\label{aform10001}{}
\end{eqnarray}
which in the classical limit correspond to the following Poisson brackets $\{X_{1},X_{2}\}=-\theta,$
$\{X_{1},P_{1}\}=\{X_{2},P_{2}\}=1+\gamma,$ $\{P_{1},P_{2}\}=\eta.$  From (\ref{aform101})-(\ref{aform10001}) follows that algebra (\ref{2dform101})-(\ref{2dform10001}) is not invariant upon time reversal.

Note that results (\ref{omega1}), (\ref{period1}) can be obtained,  taking into account that the motion in opposite direction corresponds to the time reversal transformation and upon time reversal one has (\ref{aform101})-(\ref{aform10001}). Therefore,  expressions for $\omega^{\prime}$, $T^{\prime}$ (\ref{omega1}), (\ref{period1}) can be found changing the signs of parameters of noncommutativity in (\ref{omega}), (\ref{period}) (changing $\theta$ to $-\theta$ and $\eta$ to $-\eta$). We also have that changing  signs of parameters of noncommutativity in $P^{\prime}_0$ (\ref{pp}) (changing $\theta$ to $-\theta$ and $\eta$ to $-\eta$ ) one obtains $-P_0$ (\ref{p00}).

\section{Recovering time reversal and rotational symmetries in noncommutative phase space}

To recover the time reversal and rotational symmetries in noncommutative phase space we consider idea to construct tensors of noncommutativity  involving additional coordinates and additional momenta.  On the basis of studies presented in the previous section, we can conclude that in order to preserve the time reversal symmetry the tensors $\theta_{ij}$, $\eta_{ij}$ have to transform under the time reversal as
\begin{eqnarray}
\theta_{ij}\rightarrow-\theta_{ij}, \ \ \eta_{ij}\rightarrow-\eta_{ij}.\label{ttrans}
\end{eqnarray}

Expressions for the tensors of noncommutativity on which (\ref{ttrans}) hold from the view of simplicity can be written as follows
 \begin{eqnarray}
\theta_{ij}=\frac{c_{\theta}}{\hbar}\sum_k\varepsilon_{ijk}{p}^a_{k},\label{form130}\\
\eta_{ij}=\frac{c_{\eta}}{\hbar}\sum_k\varepsilon_{ijk}{p}^b_{k},\label{for130}
\end{eqnarray}
where $c_{\theta}$, $c_{\eta}$ are constants,  ${p}^a_i$, ${p}^b_i$ are additional momenta.

To preserve the rotational symmetry additional coordinates ${a}_i$, ${b}_i$  and momenta ${p}^a_i$, ${p}^b_i$ conjugate of them  are supposed to be  governed by a rotationally symmetric systems. For simplicity the systems are considered to be harmonic oscillators
\begin{eqnarray}
H^a_{osc}=\frac{({\bf p}^{a})^{2}}{2m_{osc}}+\frac{m_{osc}\omega^2_{osc}{\bf a}^{2}}{2},\label{osca}\\
H^b_{osc}=\frac{({\bf p}^{b})^{2}}{2m_{osc}}+\frac{m_{osc}\omega^2_{osc}{\bf b}^{2}}{2},\label{oscb}
\end{eqnarray}
with  $\sqrt{\hbar}/\sqrt{{m_{osc}\omega_{osc}}}=l_{P}$ and very large frequency $\omega_{osc}$. So, the distance between the energy levels is large and harmonic oscillators put into the ground states remain in the states \cite{GnatenkoIJMPA17}. So, we propose the following noncommutative algebra
\begin{eqnarray}
[X_{i},X_{j}]=ic_{\theta}\sum_k\varepsilon_{ijk}{p}^a_{k},\label{for101}\\{}
[X_{i},P_{j}]=i\hbar\left(\delta_{ij}+\frac{c_{\theta}c_{\eta}}{4\hbar^2}({\bf {p}}^a\cdot{\bf {p}}^b)\delta_{ij}-\frac{c_{\theta}c_{\eta}}{4\hbar^2}p^{a}_j{p}^{b}_i\right),\label{for1001}\\{}
[P_{i},P_{j}]=i{c_{\eta}}\sum_k\varepsilon_{ijk}{p}^{b}_{k}.{}\label{for10001}
\end{eqnarray}
here we take into account (\ref{form130}), (\ref{for130}) and consider $\gamma_{ij}=\sum_k \theta_{ik}\eta_{jk}/4$, as was considered in \cite{Bertolami}.

Additional coordinates and additional momenta are supposed to satisfy the ordinary commutation relations
\begin{eqnarray}
[{a}_{i},{a}_{j}]=[{b}_{i},{b}_{j}]=[{a}_{i},{b}_{j}]=[{p}^{a}_{i},{p}^{a}_{j}]=[{p}^{b}_{i},{p}^{b}_{j}]=[{p}^{a}_{i},{p}^{b}_{j}]=0,{}\\{} [{a}_{i},{p}^{a}_{j}]=[{b}_{i},{p}^{b}_{j}]=i\hbar\delta_{ij},{}\\{}
[{a}_{i},{p}^{b}_{j}]=[{b}_{i},{p}^{a}_{j}]=0.{}
\end{eqnarray}
Also, relation
\begin{eqnarray}
[{a}_{i},X_{j}]=[{a}_{i},P_{j}]=[{p}^{b}_{i},X_{j}]=[{p}^{b}_{i},P_{j}]=0
\end{eqnarray}
hold. So, operators $X_{i}$, $P_{i}$ and $\theta_{ij}$, $\eta_{ij}$ satisfy the same commutation relations as in the case  of noncommutative phase space of canonical type (\ref{form101})-(\ref{form10001}).  We have $[\theta_{ij}, X_k]=[\theta_{ij}, P_k]=[\eta_{ij}, X_k]=[\eta_{ij}, P_k]=[\gamma_{ij}, X_k]=[\gamma_{ij}, P_k]=0$.  In this sense noncommutative algebra (\ref{for101})-(\ref{for10001}) is equivalent to (\ref{form101})-(\ref{form10001}).

The coordinates and momenta  ${a}_i$, ${b}_i$, ${p}^a_i$, ${p}^b_i$  can be treated as internal coordinates and momenta of a particle. Quantum fluctuations of these coordinates lead effectively to a non-point-like particle with size of the order of the Planck scale.

Because of involving of additional coordinates and additional momenta  one has to consider the total Hamiltonian defined as
\begin{eqnarray}
H=H_s+H^a_{osc}+H^b_{osc},\label{total}
\end{eqnarray}
where $H_s$ is Hamiltonian of a system under consideration and $H^a_{osc}$, $H^b_{osc}$ are given by (\ref{osca}), (\ref{oscb}).
 Taking into account that  coordinates and  momenta upon time reversal transform as
$X_{i}\rightarrow X_{i},$ $P_{i}\rightarrow -P_{i}$, $a_{i}\rightarrow a_{i},$ $p^a_{i}\rightarrow -p^a_{i}$, $b_{i}\rightarrow b_{i},$ $p^b_{i}\rightarrow -p^b_{i}$ and the time reversal operation involves complex conjugation, one obtains that  algebra (\ref{for101})-(\ref{for10001}) and Hamiltonian (\ref{total}) are invariant under the time reversal. So, the time reversal symmetry is preserved in a space with (\ref{for101})-(\ref{for10001}).

We would like to note here that because of invariance of noncommutative algebra (\ref{for101})-(\ref{for10001}) on time reversal, independently of representation one can find that upon time reversal $X_{i}\rightarrow X_{i},$ $P_{i}\rightarrow -P_{i}$. For example, the coordinates and momenta which satisfy (\ref{for101})-(\ref{for10001}) can be represented as follows
\begin{eqnarray}
X_{i}=x_{i}+\frac{c_{\theta}}{2\hbar}[{\bf p}^a\times{\bf p}]_i,\label{repx0}\\
P_{i}=p_{i}-\frac{c_{\eta}}{2\hbar}[{\bf x}\times{{\bf p}^b}]_i,\label{repp0}
\end{eqnarray}
with $x_i$, $p_i$ satisfying the ordinary commutation relations $[x_i,x_j]=[p_i,p_j]=0, [x_i,p_j]=i\hbar\delta_{ij}$.
After time reversal one has  $x_i\rightarrow x_i$, $p_i\rightarrow -p_i$, $p^a_i\rightarrow-p^a_i$, $p^b_i\rightarrow-p^b_i$ and taking into account (\ref{repx0}), (\ref{repp0}) noncommutative coordinates and noncommutative momenta transform as
$X_{i}\rightarrow X_{i}$, $P_{i}\rightarrow -P_{i}$. We would like to mention here that in the case of noncommutative algebra of canonical type the transformation of noncommutative coordinates and noncommutative momenta upon time reversal depends on their representation (see Appendix). This is a consequence of non-invariance of noncommutative algebra of canonical type on the time reversal.

Besides time-reversal invariance the algebra (\ref{for101})-(\ref{for10001}) is rotationally invariant. After rotation  $X_{i}^{\prime}=U(\varphi)X_{i}U^{+}(\varphi)$, $P_{i}^{\prime}=U(\varphi)P_{i}U^{+}(\varphi)$ $a_{i}^{\prime}=U(\varphi)a_{i}U^{+}(\varphi)$,  $p^{b\prime}_{i}=U(\varphi)p^b_{i}U^{+}(\varphi)$
the commutation relations (\ref{for101})-(\ref{for10001}) remain the same
\begin{eqnarray}
[X^{\prime}_{i},X^{\prime}_{j}]=ic_{\theta}\sum_k\varepsilon_{ijk}{p}^{a\prime}_{k},\label{fo101}\\{}
[X^{\prime}_{i},P^{\prime}_{j}]=i\hbar\left(\delta_{ij}+\frac{c_{\theta}c_{\eta}}{4\hbar}({\bf {p}}^{a\prime}\cdot{\bf {p}}^{b\prime})\delta_{ij}-\frac{c_{\theta}c_{\eta}}{4\hbar}p^{a\prime}_j{p}^{b\prime}_i\right),\label{fo1001}\\{}
[P^{\prime}_{i},P^{\prime}_{j}]=i{c_{\eta}}\sum_k\varepsilon_{ijk}{p}^{b\prime}_{k}.{}\label{fo10001}
\end{eqnarray}
Operator of rotation has the form $U(\varphi)=\exp(i\varphi({\bf n}\cdot{\bf L^t})/\hbar)$ with ${\bf L^t}=[{\bf x}\times{\bf p}]+[{\bf{a}}\times{\bf {p}}^{a}]+[{\bf{b}}\times{\bf { p}}^{b}]$, $U^{+}(\varphi)=\exp(-i\varphi({\bf n}\cdot{\bf L^t})/\hbar)$  \cite{GnatenkoIJMPA17}.

So, noncommutative algebra (\ref{for101})-(\ref{for10001}) is rotationally and time-reversal invariant and is equivalent to noncommutative algebra of canonical type.
Note that the proposed algebra is consistent. The Jacobi identity is satisfied and can be easily
checked for all possible triplets of operators because of explicit representation (\ref{repx0}), (\ref{repp0}).

We would like to mention that in our previous paper \cite{GnatenkoIJMPA17} in order to preserve rotational symmetry we proposed noncommutative algebra (\ref{for101})-(\ref{for10001}) with tensors of noncommutativity defined as
 \begin{eqnarray}
\theta_{ij}=\frac{l_0}{\hbar}\sum_k\varepsilon_{ijk}{a}_{k}, \label{f130}\\
\eta_{ij}=\frac{p_0}{\hbar}\sum_k\varepsilon_{ijk}{p}^b_{k}.\label{ff130}
\end{eqnarray}
with $l_0$, $p_0$ being constants and $a_k$, $p^b_k$ being additional coordinates and momenta governed by harmonic oscillators.  In the case when $\theta_{ij}$  is defined as (\ref{f130}) the commutation relations (\ref{form101}) are not invariant under the time reversal. Upon time reversal one has
\begin{eqnarray}
[X_{i},X_{j}]=-i{l_0}\sum_k\varepsilon_{ijk}{a}_{k}=-i\hbar\theta_{ij}\label{atr}
\end{eqnarray}
We would like also to note here that instead of examining the total Hamiltonian (\ref{total}), one can study
an effective Hamiltonian
\begin{eqnarray}
H_0=\langle H_s\rangle_{ab}+H^a_{osc}+H^b_{osc},\label{2h0}
\end{eqnarray}
 up to the second order in
\begin{eqnarray}
\Delta H= H-H_0=H_s-\langle H_s\rangle_{ab}.
\end{eqnarray}
This is because the corrections to the energy levels of the total Hamiltonian $H$ (\ref{total}) caused by terms $\Delta H$ vanish up to the second order in the perturbation theory \cite{GnatenkoIJMPA18}. Here notation $\langle...\rangle_{ab}$ is used for averaging over degrees of freedom of harmonic oscillators $H^a_{osc}$ $H^b_{osc}$ in the ground states
 \begin{eqnarray}
 \langle...\rangle_{ab}=\langle\psi^{a}_{0,0,0}\psi^{b}_{0,0,0}|...|\psi^{a}_{0,0,0}\psi^{b}_{0,0,0}\rangle
  \end{eqnarray}
$\psi^{a}_{0,0,0}$, $\psi^{b}_{0,0,0}$ are eigenstates of $H^a_{osc}$, $H^b_{osc}$.
Note that $H_0$ does not contain terms linear over parameters of noncommutativity. After averaging over $\psi^{a}_{0,0,0}$, $\psi^{b}_{0,0,0}$ the terms in Hamiltonian $H_s$ in the first order in $\theta_{ij}$, $\eta_{ij}$ vanish because of $\langle a_i\rangle_{ab}=\langle p^b_i\rangle_{ab}=0$. The Hamiltonian  $H_0$ depends only on
\begin{eqnarray}
\langle\theta^2_i\rangle=\frac{l_0^2}{\hbar^2}\langle\psi^{a}_{0,0,0}| {a}^2_i|\psi^{a}_{0,0,0}\rangle=\frac{l_0^2l_P^2}{2\hbar^2}=\frac{\langle\theta^2\rangle}{3},\label{thetar2}\\
\langle\eta^2_i\rangle= \frac{p_0^2}{\hbar^2}\langle\psi^{b}_{0,0,0}| ({p}^{b}_i)^2|\psi^{b}_{0,0,0}\rangle=\frac{p_0^2}{2\hbar^2l_P^2}=\frac{\langle\eta^2\rangle}{3},\label{etar2}
\end{eqnarray}
 Therefore, $H_0$ is invariant on replacement of $\theta_{ij}$ by $-\theta_{ij}$  and taking into account (\ref{atr}) is invariant under the time reversal transformation. Note that this statement holds for different definitions of the tensors of noncommutativity on which $\langle \theta_i\rangle_{ab}=\langle \eta_i\rangle_{ab}=0$.
We would like also to mention that Hamiltonian $H_0$ depends on the mean values $\langle\theta^2\rangle$, $\langle\eta^2\rangle$ and does not depend explicitly on the way of definition of tensors of noncommutativity $\theta_{ij}$, $\eta_{ij}$ and on the rotationally-invariant system on which  $a_i$, $b_i$  $p^a_i$, $p^b_i$ are governed by. So,  independently of definition of the tensors of noncommutativity (the only one condition has to be satisfied $\langle \theta_{ij}\rangle_{ab}=\langle \eta_{ij}\rangle_{ab}=0$) the effective Hamiltonian $H_0$ is invariant upon time reversal.

 So, idea to define tensors of noncommutativity, introducing additional coordinates and additional momenta, gives a possibility to construct noncommutative algebra which is rotationally-invariant, invariant under the time reversal transformation and equivalent to noncommutative algebra of canonical type.

\section{Conclusions}

Time reversal symmetry has been studied in a space with noncommutativity of coordinates and noncommutativity of momenta of canonical type. It has been  shown that noncommutative algebra (\ref{2dform101})-(\ref{2dform10001}) is not time reversal invariant. Upon time reversal one obtains noncommutative algebra with opposite signs of parameters of noncommutativity   (\ref{aform101}), (\ref{aform10001}). We have also  concluded that because of non-invariance of algebra (\ref{2dform101})-(\ref{2dform10001}), transformations for noncommutative coordinates and noncommutative momenta upon time reversal depend on their representation (\ref{t01010})-(\ref{t010122}).

Circular motion has been examined in noncommutative phase space as an evident example  for studying the time reversal symmetry breaking. Frequency and period of the motion have been found exactly in noncommutative phase space of canonical type. We have concluded that because of noncommutativity the frequency and the period of the circular motion depends on its direction (\ref{omega}), (\ref{period}), (\ref{omega1}), (\ref{period1}).  The effect of noncommutativity on the motion in a circle of radius $R_0$ depends on its direction.

To recover the time reversal symmetry in noncommutative phase space we have considered the idea to generalize parameters of noncommutativity to a tensors. We have shown that the time reversal symmetry is preserved if  the tensors of noncommutativity transform as $\theta_{ij}\rightarrow-\theta_{ij}$, $\eta_{ij}\rightarrow-\eta_{ij}$ under time reversal. So, on the basis of this statement and from the view of simplicity we proposed the tensors of noncommutativity to be defined as (\ref{form130}), (\ref{for130}).
To construct these tensors an additional coordinates and additional momenta have been considered. For preserving the rotational symmetry the coordinates and the momenta have been supposed to be governed by a rotationally-invariant systems which for simplicity are considered to be harmonic oscillators.  As a result we propose noncommutative algebra (\ref{for101})-(\ref{for10001}) which is rotationally invariant, time reversal invariant besides it is equivalent to noncommutative algebra of canonical type.

\section*{Acknowledgments}
 This work was partly supported by the projects $\Phi\Phi$-63Hp (No. 0117U007190), $\Phi\Phi$-30$\Phi$ (No. 0116U001539) from the Ministry of Education and Science of Ukraine.

\section*{Appendix}

Because of non-invariance of noncommutative algebra of canonical type (\ref{2dform101})-(\ref{2dform10001}) under the time reversal, the transformation of $X_i$, $P_i$ upon time reversal depends on representation. It is known that the noncommutative coordinates and  noncommutative momenta which satisfy  (\ref{2dform101})-(\ref{2dform10001}) can be represented by coordinates and momenta $x_i$, $p_i$ with
\begin{eqnarray}
[x_{i},x_{j}]=0,\label{or}\\{}
[x_{i},p_{j}]=i\hbar\delta_{ij},\\{}
[p_{i},p_{j}]=0.\label{or1}
\end{eqnarray}
Namely, for coordinates and momenta which satisfy (\ref{2dform101})-(\ref{2dform10001}) we can write
\begin{eqnarray}
X_{1}=\varepsilon\left(x_{1}-\theta'_1{p}_{2}\right),\label{for01010}\\
X_{2}=\varepsilon\left(x_{2}+\theta'_2{p}_{1}\right),\\
P_{1}=\varepsilon\left(p_{1}+\eta'_1{x}_{2}\right),\\
P_{2}=\varepsilon\left(p_{2}-\eta'_2{x}_{1}\right),\label{for010122}
\end{eqnarray}
with $\varepsilon$, $\theta'_1$, $\theta'_2$, $\eta'_2$, $\eta'_2$ being constants. Upon time reversal, considering transformations $x_i\rightarrow x_i$, $p_i\rightarrow -p_i$, we have
\begin{eqnarray}
X_{1}\rightarrow X^{\prime}_{1}=\varepsilon\left(x_{1}+\theta'_1{p}_{2}\right),\label{t01010}\\
X_{2}\rightarrow X^{\prime}_{2}=\varepsilon\left(x_{2}-\theta'_2{p}_{1}\right),\\
P_{1}\rightarrow -P^{\prime}_{1}=\varepsilon\left(-p_{1}+\eta'_1{x}_{2}\right),\\
P_{2}\rightarrow -P^{\prime}_{2}=\varepsilon\left(-p_{2}-\eta'_2{x}_{1}\right).\label{t010122}
\end{eqnarray}
 So, transformations (\ref{t01010})-(\ref{t010122}) depend on the parameters $\varepsilon$, $\theta'_1$, $\theta'_2$, $\eta'_2$, $\eta'_2$, therefore they depend on the representation.

  Parameters $\varepsilon$, $\theta'_1$, $\theta'_2$, $\eta'_2$, $\eta'_2$, can be chosen in different ways.
Taking into account (\ref{or})-(\ref{or1}), (\ref{for01010})-(\ref{for010122}) one has
  \begin{eqnarray}
[X_{1},X_{2}]=i\hbar\varepsilon^2(\theta'_1+\theta'_2),\label{m101}\\{}
[X_{1},P_{1}]=i\hbar\varepsilon^2(1+\theta'_1\eta'_1)\\{}
[X_{2},P_{2}]=i\hbar\varepsilon^2(1+\theta'_2\eta'_2),\\{}
[P_{1},P_{2}]=i\hbar\varepsilon^2(\eta'_1+\eta'_2).\label{m10001}{}
\end{eqnarray}
On the basis of comparison of (\ref{m101})-(\ref{m10001}) with (\ref{2dform101})-(\ref{2dform10001}) we can write the following equations
\begin{eqnarray}
\varepsilon^2=1, \ \ \theta'_1\eta'_1=\theta'_2\eta'_2=\gamma, \label{eq1}\\{}
\theta'_1+\theta'_2=\theta,\\
\eta'_1+\eta'_2=\eta,\label{eq2}
\end{eqnarray}
from which we obtain
\begin{eqnarray}
\theta'_1=\frac{1}{2}\left({\theta\pm\sqrt{\theta^2-4\frac{\theta\gamma}{\eta}}}\right),\label{tt}\\
\theta'_2=\frac{1}{2}\left({\theta\mp\sqrt{\theta^2-4\frac{\theta\gamma}{\eta}}}\right),\\
\eta'_1=\frac{1}{2}\left({\eta\mp\sqrt{\eta^2-4\frac{\eta\gamma}{\theta}}}\right),\\
\eta'_2=\frac{1}{2}\left({\eta\pm\sqrt{\eta^2-4\frac{\eta\gamma}{\theta}}}\right),\label{ee}
\end{eqnarray}
and $\gamma\leq\theta\eta/4$.
So, choosing the signs in (\ref{tt})-(\ref{ee}) one obtains two different representations for noncommutative coordinates and noncommutative momenta and therefore two different transformations upon time reversal (\ref{t01010})-(\ref{t010122}).

Symmetric representation with $\varepsilon=1$, $\theta'_1=\theta'_2=\theta/2$, $\eta'_1=\eta'_2=\eta/2$ is well known. In this case  coordinates and momenta $X_i$, $P_i$ satisfy (\ref{form101})-(\ref{form10001}) with $\gamma=\theta\eta/4$  \cite{Bertolami3}.

For $\gamma=0$ in (\ref{2dform1001})  the commutator of coordinates and momenta is equal to $i\hbar$  as in the ordinary space. Comparing (\ref{m101})-(\ref{m10001}) and (\ref{2dform101})-(\ref{2dform10001}) with $\gamma=0$ we can write
\begin{eqnarray}
\varepsilon^2=\frac{1}{1+\theta'_1\eta'_1},\label{eq11}\\
\theta'_1\eta'_1=\theta'_2\eta'_2,\\
\varepsilon^2(\theta'_1+\theta'_2)=\theta,\\
\varepsilon^2(\eta'_1+\eta'_2)=\eta,\label{eq22}
\end{eqnarray}
Note that we have four equations (\ref{eq11})-(\ref{eq22}) and five parameters $\varepsilon$, $\theta'_1$, $\theta'_2$, $\eta'_1$, $\eta'_2$. So, choosing one of them one can obtain different representations of coordinates and momenta which satisfy  (\ref{2dform101})-(\ref{2dform10001}) with $\gamma=0$ and different transformations (\ref{t01010})-(\ref{t010122}).
For instance, one can choose $\theta'_2=0$. As a result form (\ref{eq11})-(\ref{eq22}) one obtains $\varepsilon=1$,  $\eta'_1=0$, $\eta'_2=\eta$, $\theta'_1=\theta$ and the  representation reads
$X_{1}=x_{1}-\theta{p}_{2},$ $X_{2}=x_{2}$, $P_{1}=p_{1}$, $P_{2}=p_{2}-\eta{x}_{1}$. In this case upon time reversal the coordinate $X_2$, and momentum $P_1$  transform in the traditional way
$X_{2}\rightarrow X_{2},$
$P_{1}\rightarrow -P_{1}$. But for $X_1$, $P_1$ one has
$X_{1}\rightarrow X^{\prime}_{1}=x_{1}+\theta{p}_{2},$
$P_{2}\rightarrow -P^{\prime}_{2}=-p_{2}-\eta{x}_{1}.$

It is also possible to write two symmetric representations (\ref{for01010})-(\ref{for010122}) with parameters $\varepsilon=(1+{\theta'\eta'})^{-\frac{1}{2}}$, $\theta'_1=\theta'_2=(1\pm\sqrt{1-\theta\eta})/\eta$, $\eta'_1=\eta'_2=(1\pm\sqrt{1-\theta\eta})/\theta$  \cite{Bertolami3,GnatenkoMPLA17} which lead to different transformations under the time reversal.


\begin{thebibliography}{99}
\bibitem{Snyder} H.~Snyder, Phys. Rev. {\bf71}, 38 (1947).
\bibitem{Chaichian} M.~Chaichian, M.M.~Sheikh-Jabbari, A. Tureanu, Phys. Rev. Lett.  {\bf 86},  2716 (2001).
\bibitem{Balachandran1} A.P.~Balachandran, P.~Padmanabhan, J. High Energy Phys.  {\bf1012}, 001 (2010).
\bibitem{Geloun} J. Ben Geloun, Sunandan Gangopadhyay, F. G. Scholtz,  EPL {\bf86}, 51001 (2009).
\bibitem{Scholtz} F. G. Scholtz, L. Gouba, A. Hafver, C. M. Rohwer, J. Phys. A {\bf42}, 175303 (2009).

\bibitem{Moreno} E. F. Moreno,  Phys. Rev. D {\bf72}, 045001 (2005).
\bibitem{Galikova} V. G\'alikov\'a, P. Presnajder, J. Phys: Conf. Ser. {\bf343}, 012096 (2012).

\bibitem{Amorim} R. Amorim, Phys. Rev. Lett. {\bf101}, 081602 (2008).
\bibitem{GnatenkoPLA14} Kh.P. Gnatenko, V. M. Tkachuk,  Phys. Lett. A {\bf378}, 3509 (2014).

    \bibitem{Lukierski} M. Daszkiewicz, J. Lukierski, M. Woronowicz,  Phys. Rev. D {\bf77}, 105007 (2008).
\bibitem{Lukierski2009} M. Daszkiewicz, J. Lukierski, M. Woronowicz, J. Phys. A: Math. Theor. {\bf42}, 355201 (2009).
\bibitem{BorowiecEPL} A. Borowiec, Kumar S. Gupta, S. Meljanac,  A. Pachol,  EPL  {\bf92}, 20006 (2010).
\bibitem{Borowiec} A. Borowiec, J. Lukierski, A. Pachol, J. Phys. A: Math. Theor. {\bf47}, 405203 (2014).
\bibitem{Borowiec1} A. Borowiec, A. Pachol, SIGMA  {\bf10}, 107 (2014).
\bibitem{Kupriyanov2009}  M. Gomes, V.G. Kupriyanov,  Phys. Rev. D {\bf79}, 125011 (2009).
\bibitem{Kupriyanov} V. G. Kupriyanov, J. Phys. A: Math. Theor. {\bf46}, 245303 (2013).

\bibitem{Falomir09} H. Falomir, J. Gamboa, J. Lopez-Sarrion, F. Mendez, P.A.G. Pisani,  Phys. Lett. B
{\bf680}, 384 (2009).
\bibitem{Ferrari13} A.F. Ferrari, M. Gomes, V.G. Kupriyanov, C.A. Stechhahn, Phys. Lett. B {\bf718}, 1475 (2013).
\bibitem{Bertolami} O. Bertolami, R. Queiroz,  Phys. Lett. A {\bf375}, 4116 (2011).
\bibitem{Domingos} J. M. Domingos, Int. J Theor. Phys., {\bf18}, 213 (1979).
\bibitem{GnatenkoIJMPA17} Kh. P. Gnatenko, V. M. Tkachuk,  Int. J. Mod. Phys. A {\bf32}, 1750161 (2017).
\bibitem{GnatenkoIJMPA18} Kh. P. Gnatenko, V. M. Tkachuk,  Int. J. Mod. Phys. A {\bf33}, 1850037 (2018).
\bibitem{Bertolami3} O. Bertolami, J. G. Rosa, C. M. L. de Aragao, P. Castorina, D. Zappala,  Mod. Phys. Lett. A {\bf21}, 795 (2006).
\bibitem{GnatenkoMPLA17} Kh. P. Gnatenko,  Mod. Phys. Lett. A {\bf32},  1750166 (2017).




\end{thebibliography}
\end{document}